\documentclass[12pt,fleqn]{article}
\makeatletter \@addtoreset{equation}{section} \makeatother

\newcommand {\Fds}    {Feynman diagrams}
\newcommand {\PV}     {Pauli--Villars}

\newcommand {\CS}     {Chern--Simons}
\newcommand {\MM}     {M}                        %% 4-manifold in math mode
\newcommand {\bcs}    {boundary conditions}

\newcommand {\lhs}    {left-hand side}

\newcommand {\SM}     {Standard Model}

\newcommand {\qfth}   {quantum field theory}
\newcommand {\qfths}  {quantum field theories}

\newcommand {\YM}     {Yang--Mills}
\newcommand {\YMth}   {Yang--Mills theory}

\newcommand {\cgth}   {chiral gauge theory}
\newcommand {\cgths}  {chiral gauge theories}

\newcommand {\nacgths}{non-Abelian chiral gauge theories}

\newcommand {\rmd}    {{\rm d}}

\newcommand {\gsim}{\mathrel{\hbox{\rlap{\lower.55ex \hbox {$\sim$}}
            \kern-.3em \raise.4ex \hbox{$>$}}}}
\newcommand {\lsim}{\mathrel{\hbox{\rlap{\lower.55ex \hbox {$\sim$}}
            \kern-.3em \raise.4ex \hbox{$<$}}}}

\newcommand {\beq} {\begin{equation}}
\newcommand {\eeq} {\end{equation}}

\def\id{{\rm 1\kern-.12em
\rule{0.3pt}{1.5ex}\raisebox{0.0ex}{\rule{0.1em}{0.3pt}}}}
\def\C{{\rm\kern.24em
   \vrule width.02em
       height1.4ex depth-.05ex
   \kern-.26em C}}
\def\R  {{\rm I\kern-.15em R}}
\def\L  {{\rm I\kern-.25em L}}
\def\Z{\mbox{\sf Z\hspace{-1.8ex} \sf Z}}   % ganze Zahlen
\def\N{{\rm I\kern-.23em N}}
\begin{document}

\begin{titlepage}
\noindent Nuclear Physics B 578 (2000) 277--289               \hspace*{\fill}
hep-th/9912169   \newline
                             \hspace*{\fill}  NBI--HE--99-49 \newline
                             \hspace*{\fill}  KA--TP--20--1999
\begin{center}
\vspace{2\baselineskip} {\Large \bf A CPT anomaly}\\
\vspace{2\baselineskip} {\large F.R. K{\sc linkhamer}} \footnote{E-mail:
                          frans.klinkhamer@physik.uni-karlsruhe.de}\\
\vspace{1\baselineskip} {\it The Niels Bohr Institute, University of
Copenhagen,\\
     Blegdamsvej 17, DK--2100 Copenhagen \O, Denmark}\\
     \vspace{0.5\baselineskip}
     and\\
     \vspace{0.5\baselineskip}
{\it Institut f\"ur Theoretische Physik, Universit\"at Karlsruhe,\\
     D--76128 Karlsruhe, Germany} \footnote{Permanent address.}\\
\vspace{3\baselineskip} {\bf Abstract} \\
\end{center}
{\noindent We consider chiral gauge theories defined over a
four-dimensional spacetime manifold with a Cartesian product structure for
at least one compact spatial dimension. For a simple setup,
we calculate the effective gauge field action by integrating out %%
the chiral fermions, while maintaining gauge invariance. Due to a
combination of infrared and ultraviolet effects, there appears a CPT-odd
term in the effective gauge field action. This CPT anomaly could occur in
chiral gauge theories relevant to elementary particle physics, provided
the spacetime manifold has the appropriate topology. Two possible
applications for cosmology are discussed. } \vspace{1\baselineskip}
\begin{tabbing}
PACS \hspace{1.25em} \= : \hspace{0.25em} \=
                          11.15.-q; 11.30.Rd; 11.30.Cp; 11.30.Er
                    \\[0.5ex]
Keywords       \> : \> Gauge invariance; Chiral anomaly;
                       Lorentz noninvariance; CPT violation
\\
\end{tabbing}
\end{titlepage}

\section{Introduction}

  The CPT theorem is one of the most important results in
flat-spacetime \qfth~\nocite{L57,S64,SW64}[1--3].   %%careful
The theorem states that the combined operation (CPT) of charge conjugation
(C), parity reflection (P) and time reversal (T) is an invariance of local
relativistic \qfth, even if some of the separate invariances do not hold.
Any CPT violation is, therefore, believed to require fundamentally
different physics, for example quantum gravity \cite{W80,H82} or strings
\cite{KP91}. It may, then, come as a surprise that a particular class of
local relativistic \qfths~has given an indication of CPT violation
\cite{K98}.

  The specific theories considered in Ref. \cite{K98} are \nacgths~with one
compact spatial dimension singled out by the prescribed four-dimensional
spacetime manifold $\MM$. An example would be $SU(3)$ \YMth~\cite{YM54}
with a single triplet of left-handed Weyl fermions \cite{W29}, defined
over the flat spacetime manifold $\MM$ $=$ $\R^3 \times S^1$, which
corresponds to the usual Minkowski spacetime with one spatial coordinate
compactified to a circle. The perturbative chiral gauge anomalies of this
theory
\nocite{A69,BJ69,B69,GJ72} [10--13]  %%careful
can be cancelled by the introduction of an octet of elementary
pseudoscalar fields with the standard gauged Wess-Zumino term in the
action \cite{WZ71, W83}. There remains a nonperturbative $SU(3)$ gauge
anomaly \cite{KR97}, which is similar to, but not the same as, the Witten
$SU(2)$ gauge anomaly \cite{W82}. In this case, however, there exists a
local counterterm for the action which restores $SU(3)$ gauge invariance,
but at the price of Lorentz noninvariance and CPT violation \cite{K98}. In
other words, the remaining non-Abelian chiral gauge anomaly is transmuted
into a CPT anomaly. (The situation is analogous to that of certain
three-dimensional non-Abelian gauge theories with massless fermions, where
gauge invariance is restored at the price of P and T violation and the
non-Abelian gauge anomaly is transmuted into the so-called parity anomaly
\nocite{R84,NS83,AW84,ADM85,P87,CL89}  [18--23].)   %%careful

  The particular counterterm presented in Ref. \cite{K98} is the spacetime
integral of a \CS~density \cite{CS74} involving three of the four gauge
potentials. (The precise definition will be given later.) Such a term
obviously violates local Lorentz invariance.  Also, the integrand of the
counterterm is CPT-odd, whereas the standard \YM~action density is
CPT-even. This Lorentz and CPT noninvariance would show up, to first
order, as a direction-dependent, but wavelength-independent, rotation of
the linear polarization of a plane wave of gauge fields traveling \emph{in
vacuo} \cite{CFJ90}.

  We have obtained some heuristic arguments of why the counterterm must violate
Lorentz and CPT invariance, but the uniqueness of the counterterm has not
been established. If, on the other hand, the non-Abelian chiral gauge
anomaly is really as discussed above, then the effective gauge field
action due to the chiral fermions, formulated and regularized in a
gauge-invariant manner, must already exhibit some sign of CPT violation
(and Lorentz noninvariance). It is the goal of the present paper to
establish this CPT violation. Maintaining chiral gauge invariance, we will
find for spacetime manifolds with the appropriate Cartesian product
structure a CPT-violating term in  the effective gauge field action which
is precisely equal to the counterterm presented in Ref. \cite{K98}. This
CPT-violating term can appear in anomaly-free chiral gauge theories but
not in  vectorlike gauge theories such as quantum electrodynamics.

  The outline of this paper is as follows. In Section 2, we give the
setup of the problem and establish our notation. In Section 3, we choose
some simple background gauge potentials and find a \CS~term in the
effective gauge field action.  The calculation applies to both Abelian and
non-Abelian chiral gauge groups, provided the theory is free of chiral
gauge anomalies. In Section 4, we show that the \CS~term found violates
CPT. In other words, these particular chiral gauge field theories have a
CPT anomaly if gauge invariance is maintained. In Section 5, we give some
generalizations of our basic result. Also, we exhibit a class of
\cgths~which necessarily have the CPT a\-no\-ma\-ly, as long as the
spacetime manifold has the appropriate topology. Remarkably, the so-called
\SM~of elementary particle physics (with three families of quarks and
leptons) can be embedded in some of these anomalous theories. In Section
6, finally, we present some remarks on how the CPT theorem is circumvented
and discuss two possible applications of the CPT anomaly.

\section{Setup}

  For definiteness, we take spacetime to be the flat Euclidean manifold
\begin{equation}\label{eq:MM}
  \MM = \R^3 \times S^1 \; ,
\end{equation}
with Cartesian coordinates $x^m \in \R^3$, $m=1$, 2, 3, and $x^4 \in S^1$.
At the end of the calculation, we can make the Wick rotation \cite{IZ80}
from Euclidean to Lorentzian metric signature, with $x^4$ corresponding to
a compact spatial coordinate and $x^1$, say, to the time coordinate. The
length of the circle in the 4-direction is denoted by $L$. Throughout this
paper, Latin indices $k$, $l$, $m$, etc. run over the coordinate labels 1,
2, 3, and Greek indices $\kappa$, $\lambda$, $\mu$, etc. over 1, 2, 3, 4.
Repeated coordinate (and internal) indices are summed over. Also, natural
units are used for which $c$ $=$ $\hbar$ $=$ $k$ $=$ $1$.

We will first consider non-Abelian chiral gauge theories with a
\emph{single} irreducible representation of massless left-handed fermions.
Specifically, we take the standard chiral \YMth~\cite{YM54,W29} with gauge
group $G$ $=$ $SO(10)$ and left-handed Weyl fermions in the complex
representation $R_{\,L}$ $=$ $\mathbf{16}$. (This particular model may
have some relevance for elementary particle physics, as part of a
so-called grand-unified theory. See Refs. \cite{GG74,Z82} and references
therein.) The left-handed fermion field is then $\psi_{L \alpha i}(x)$,
with a spinor index $\alpha=1$, 2, and an internal symmetry index $i=1$,
$\ldots$ , $16$. The gauge potentials are $A_\mu(x)$ $\equiv$ $e\,
A_\mu^a(x) \,T^a$, with $e$ the gauge coupling constant and $T^a$, $a=1$,
$\ldots$ , $45$, the anti-Hermitian generators of the Lie group $SO(10)$
in the representation chosen, normalized by ${\rm tr}\, (T^a T^b)$ $=$ $-
\, \frac{1}{2}\, \delta^{ab}$.
The fermion and gauge fields are periodic in $x^4$, with period $L$.

  In this paper, we are interested in the effective gauge field action
obtained from integrating out the chiral fermions, while maintaining gauge
invariance. Formally, we have the following functional integral
\cite{IZ80}:
\begin{equation}\label{eq:GammaW}
\exp\left\{-\Gamma_{\rm W}[A\,]\right\} =
          \int \mathcal{D}\psi_L^\dagger\, \mathcal{D}\psi_L\;
          \exp\left\{-{\rm I}_{\rm \,W}
          \left[\,\psi_L^\dagger,\psi_L,A\,\right]\right\} \; ,
\end{equation}
for the Euclidean Weyl action
\begin{eqnarray}\label{eq:IW}
{\rm I}_{\rm \,W}\left[\,\psi_L^\dagger,\psi_L,A\,\right] &=&
               \int_{\MM} \rmd^4 x \; \psi_L^\dagger \;  i \sigma_{-}^\mu \,
                   \left( \partial_\mu + A_\mu\right) \, \psi_L          \;,
\end{eqnarray}
with $ \sigma_\pm^\mu \equiv (\pm \,i\sigma^m, \id\,)$ defined in terms of
the $2 \times 2$ Pauli matrices $\sigma^m$ and the $2 \times 2$ identity
matrix $\id\,$. The $SO(10)$ chiral gauge theory is anomaly free
\cite{B69,GJ72,KR97,W82} and the effective gauge field action is invariant
under local gauge transformations,
\begin{equation}\label{eq:GammaWinvar}
\Gamma_{\rm \,W}\,[\,g\,(A+\rmd)\,g^{-1}\,] =  \Gamma_{\rm \,W}\,[A\,]\; ,
                                               \quad g(x) \in G \; ,
\end{equation}
with $\rmd$ the exterior derivative for differential forms ($\rmd g$
$\equiv$ $\partial g/\partial x^\mu$ $\rmd x^\mu$) and $A$ $\equiv$ $A_\mu
\, \rmd x^\mu$ a one-form taking values in the Lie algebra (here, in the
defining representation).

If the chiral gauge theory considered is not anomaly free (for example,
the theory mentioned in the Introduction, with $G$ $=$ $SU(3)$ and
$R_{\,L}$ $=$ $\mathbf{3}$), then the theory has to be modified in order
to make it gauge invariant. One way to restore gauge invariance is by
averaging over the gauge orbits,
\begin{equation}\label{eq:Gamma}
\exp\left\{-\Gamma\,[A\,]\right\} \equiv
          \int \mathcal{D}h \; \exp
          \left\{-\,\Gamma_{\rm \,W}\,[\,h\,(A+\rmd)\,h^{-1}\,]
          \right\}\;.
\end{equation}
But the interpretation of the resulting theory with the dimensionless
variables $h(x)$ $\in$ $G$ is not entirely clear \cite{FS86}. Another way
to restore gauge invariance is by introducing further fermions, which
cancel the chiral anomalies of the original fermions \cite{GJ72}. In
Section 5, we will discuss some of these theories with reducible fermion
representations. All of these complications are, however, not necessary
for the anomaly-free chiral gauge theory considered here, which has the
gauge group $G$ $=$ $SO(10)$ and the fermion representation $R_{\,L}$ $=$
$\mathbf{16}$.

At this point, there is no need to be explicit about the regularization of
the effective gauge field action $\Gamma_{\rm \,W}\,[A\,]$. One possible
regularization would be the introduction of a spacetime lattice cutoff,
which (temporarily?) sacrifices Lorentz invariance but keeps the gauge and
chiral invariances intact. (See Refs. \cite{N98,L99} and references
therein.) This last condition on the regularization method is important,
since we intend to look for symmetry violations being forced upon us by
maintaining exact gauge invariance in a theory with genuine chiral
fermions.

\section{Calculation}

  As discussed in the Introduction, our goal is to establish
the presence of a CPT-violating term in the effective gauge field action
for the theory defined in Section 2. The strategy is to simplify the
calculation as much as possible. We, therefore, take the case of
$x^4$-independent $SO(10)$ gauge potentials, with the one gauge potential
corresponding to the special direction (here, $x^4 \in S^1$ for the
Euclidean spacetime  manifold $\MM$ $=$ $\R^3 \times S^1$) vanishing
altogether, \beq\label{eq:Atilde} A_m(\vec{x},x^4) =
\tilde{A}_m(\vec{x})\; ,  \quad A_4(\vec{x},x^4) = \tilde{A}_4(\vec{x})= 0
\; . \eeq Also, the gauge potentials considered vanish on the boundary of
a ball $B^3$ embedded in $\R^3$, and outside of it,
\begin{equation}\label{eq:ball}
\tilde{A}_m(\vec{x})=0 \quad {\rm for} \quad |\vec{x}| \geq R \; ,
\end{equation}
with $R$ a fixed radius which can be taken to infinity at the end of the
calculation.

The left-handed fermion field $\psi_L$ in the complex representation
$R_{\,L}$ $=$ $\mathbf{16}$ of $SO(10)$ and the independent fermion field
$\psi_L^\dagger$ in the conjugate representation can be expanded in
Fourier modes
\begin{eqnarray}\label{eq:psiL}
\psi_L(\vec{x},x^4) &=& \sum_{n=-\infty}^{\infty}\,
                        e^{+2\pi i n x^4/L}\; \xi_n(\vec{x})\, , \nonumber\\
\psi_L^\dagger(\vec{x},x^4) &=& \sum_{n=-\infty}^{\infty}\,
                        e^{-2\pi i n x^4/L}\; \xi_n^\dagger(\vec{x})\, .
\end{eqnarray}
The Weyl action (\ref{eq:IW}) for the gauge potentials (\ref{eq:Atilde})
then becomes
\begin{eqnarray}\label{eq:IWeyl}
{\rm I}_{\rm \,W} &=&
               \sum_{n=-\infty}^{\infty}\,\int_{\R^3} \rmd^3 x \;
               L\; \,\xi_n^\dagger
               \left(\,\sigma^m \,(\partial_m+\tilde{A}_m)-2\pi n/L\,\right)
               \xi_n\; .
\end{eqnarray}
Redefining the two independent sets of spinor fields
\begin{equation}\label{eq:chi}
\chi_n(\vec{x}) \equiv i \, L\; \xi_n(\vec{x}) \; , \quad
\chi_n^\dagger(\vec{x}) \equiv  \xi_n^\dagger(\vec{x}) \; ,
\end{equation}
the action reads
\begin{eqnarray}\label{eq:sumI3}
{\rm I}_{\rm \,W} &=& \sum_{n=-\infty}^{\infty}\,\int_{\R^3} \rmd^3 x \;
               \chi_n^\dagger
               \left(-\,i\,\sigma^m(\partial_m+\tilde{A}_m)+i\,2\pi n/L\,\right)
               \chi_n \nonumber\\
        &\equiv& \sum_{n=-\infty}^{\infty}\,
                {\rm I}_{\,3} \left[\,\chi_n^\dagger,\chi_n,\tilde{A}\,\right]\; .
\end{eqnarray}
We have thus obtained an infinite set of three-dimensional Euclidean Dirac
fields $\chi_n(\vec{x})$ with masses $2 \pi n/L$, all of which interact
with the \emph{same} three-dimensional gauge potentials
$\tilde{A}_m(\vec{x})$. (This is, of course, reminiscent of Kaluza-Klein
theory, which reduces five spacetime dimensions to four. See Refs.
\cite{Z82,P58} and references therein.)

For the special gauge potentials (\ref{eq:Atilde}), the effective action
(\ref{eq:GammaW}) now factorizes to
\begin{equation}\label{eq:GammaWfactor}
\exp\left\{-\Gamma_{\rm W}[\tilde{A}\,]\right\} \propto
                       \prod_{n=-\infty}^{\infty}\left(\;
                       \int \mathcal{D}\chi_n^\dagger\, \mathcal{D}\chi_n\;
                       \exp\left\{-{\rm I}_{\,3}
                       \left[\,\chi_n^\dagger,\chi_n,\tilde{A}\,\right]
                       \right\}\;\right) \; ,
\end{equation}
with the three-dimensional action ${\rm I}_{\,3}$ as defined in
(\ref{eq:sumI3}). Each factor in (\ref{eq:GammaWfactor}) can be
regularized separately by the introduction of appropriate
\emph{three-dimensional} \PV~fields \cite{IZ80,PV49}. This ultraviolet
regularization preserves the restricted gauge invariance
\begin{equation}\label{eq:3dgaugetr}
\chi_n \rightarrow  U_r(\tilde{g}) \;\chi_n\;, \quad \tilde{A}_m^{\,(r)}
\rightarrow  U_r(\tilde{g})
                     \left( \tilde{A}_m^{\,(r)} + \partial_m \right)
                      U_r^{-1}(\tilde{g})\; ,
                     \quad \tilde{g}(\vec{x}) \in G \; ,
\end{equation}
with $U_r$ the appropriate unitary representation for the fermions (here,
$r$ $=$ $\mathbf{16}$ and $G$ $=$ $SO(10)$) and gauge functions
$\tilde{g}(\vec{x})$ $=$ $\id\,$ for $|\vec{x}|$ $\geq$ $R$. Even though
this is not the full gauge invariance (\ref{eq:GammaWinvar}) of the
theory, it turns out to be sufficient for our purpose (see Section 4).

In addition to the ultraviolet divergences in the separate factors of
(\ref{eq:GammaWfactor}), which are regularized by the corresponding
three-dimensional \PV~fields, there are also infrared divergences in the
$n = 0$ factor. These infrared divergences can be regularized by imposing
antiperiodic \bcs~for the Dirac (and \PV) fields on the surface of the
ball $B^3$, where the gauge potentials (\ref{eq:ball}) vanish.

The massive \PV~regulator fields for the $n=0$ factor of
(\ref{eq:GammaWfactor}), viewed as $x^4$-in\-de\-pen\-dent
four-dimensional fields, introduce a breaking of Lorentz and CPT
invariance in the four-dimensional con\-text. This breaking will show up
later as a finite remnant in the effective gauge field action.
(Preliminary results seem to indicate that this is also the case for the
lattice regularization mentioned in the last paragraph of Section 2.)

For the present calculation, it is sufficient to introduce for each
(anticommuting) field  $\chi_n(\vec{x})$ with mass $M_n$ $\equiv$ $2\pi
n/L\,$ a single (commuting) \PV~field $\phi_n(\vec{x})$ with mass
$\Lambda_0$ for $n=0$ and $\Lambda_n$ $\equiv$  $M_n$ $+$ ${\rm
sign}(n)\,\Lambda$ for $n\neq 0$, where $\Lambda$ is taken to be positive.
Formally, this gives for (\ref{eq:GammaWfactor}) the following product:
\begin{equation}\label{eq:GammaWproduct}
\exp\left\{-\Gamma_{\rm W}[\tilde{A}\,]\right\} \propto \,
     \prod_{k} \, \frac{\lambda_k}{\lambda_k + i\Lambda_0} \,
     \left(\, \prod_{l=1}^{\infty}
     \frac{\lambda_k^2 + M_l^2}{\lambda_k^2 + (M_l+\Lambda)^2}\,\right)\; ,
\end{equation}
in terms of the real eigenvalues $\lambda_k$ of the massless
three-dimensional Dirac operator $-\,i\,\sigma^m$
$(\partial_m+\tilde{A}_m)$. The factors in (\ref{eq:GammaWfactor}) with
$n=\pm \,l$, for $l>0$, thus combine to give a \emph{real} contribution to
the effective gauge field action $\Gamma_{\rm W}[\tilde{A}\,]$. Moreover,
it is clear that the spectral flow \cite{R84,ADM85} of the full
three-dimensional Dirac operator  as given in (\ref{eq:sumI3}) can occur
only in the $n = 0$ sector (there is a mass gap for $n \ne 0$), and that
the potential non-Abelian gauge anomaly \cite{KR97}, which shows up in the
imaginary part of $\Gamma_{\rm W}[\tilde{A}\,]$, resides there.

The imaginary part of the effective gauge field action for massless
three-dimensional Dirac fermions, with \PV~regularization to maintain
gauge invariance, has already been calculated \cite{R84,ADM85}. Revisiting
the perturbative calculation, we have for the $n = 0$ sector of our
non-Abelian $SO(10)$ gauge theory the one-loop result
\begin{equation}\label{eq:GammaWn0}
\Gamma_{\rm \,W}^{\,n=0}\,[\tilde{A}\,]
       \supseteq \, i \int_{B^3}\, \rmd^3 x\; s_0 \, \pi \,
       \omega_{\rm \,CS}[\tilde{A}_1, \tilde{A}_2, \tilde{A}_3\,]\; ,
\end{equation}
in terms of a sign factor $s_0 = \pm \, 1$ whose origin will be explained
shortly and the \CS~density \cite{ADM85,CS74}
\begin{equation}\label{eq:omegaCS}
\omega_{\rm \,CS} [ A_1,A_2,A_3 \,] \equiv
       \,\frac{1}{16\, \pi^2} \;\epsilon^{klm}\; {\rm tr}
       \left( A_{kl}\, A_m - {\textstyle \frac{2}{ 3}}\,A_k\,A_l\,A_m \right)\;,
\end{equation}
with indices $k$, $l$, $m$, running over 1, 2, 3. Here, $\epsilon^{klm}$
is the completely antisymmetric Levi-Civita symbol, normalized to
$\epsilon^{123}=+1$, and $A_{kl}$ $\equiv$ $\partial_k A_l$ $-$
$\partial_l A_k$ $+$ $\left[\,A_k\, ,A_l\,\right]\,$ is the field strength
tensor for the gauge potential $A_m$ $\equiv$ $e\, A_m^a \,T^a$, with
gauge coupling constant $e$ and  anti-Hermitian Lie group generators
$T^a$, normalized by ${\rm tr}\, (T^a T^b)$ $=$ $- \, \frac{1}{2}\,
\delta^{ab}$.

The sign ambiguity $s_0$ in (\ref{eq:GammaWn0}) traces back to the
parity-violating \PV~mass $\Lambda_0$ used to regularize the ultraviolet
divergences of the three-dimensional \Fds. (Here, parity violation is
meant in the three-dimensional sense. As will become clear in the next
section, three-dimensional parity corresponds effectively to CPT in the
four-dimensional context \cite{K98}.) The factor $s_0$ in
(\ref{eq:GammaWn0}) comes, in fact, from a factor $\Lambda_0/|\Lambda_0|$
out of  the momentum integrals. The triangle dia\-gram, for example, gives
in the limit $|\Lambda_0|$ $\rightarrow$ $\infty$
\begin{equation}\label{eq:s0}
\pi^{-2}\,\int_{0}^{\infty}\,\rmd q\; 4\pi
q^2\,\Lambda_0\,(q^2+\Lambda_0^2)\, (q^2+\Lambda_0^2)^{-3} \,=\,
\Lambda_0/|\Lambda_0|  \,\equiv\, s_0  \; ,
\end{equation}
with the explicit factor $\Lambda_0\,(q^2+\Lambda_0^2)$ from the spinor
trace in the integrand on the \lhs. It is also important that the infrared
divergences of the three-dimensional Feynman diagrams without \PV~fields
are \emph{not} regularized by the introduction of a small Dirac mass
$\lambda_0$, which would again violate parity invariance, but that they
are kept under control by the antiperiodic \bcs~imposed on the fermions
(turning the momentum integrals into sums).

The essential conditions for the derivation of (\ref{eq:GammaWn0}) are
thus the requirement of gauge invariance (\ref{eq:3dgaugetr}) and the
control of infrared divergences in the $n=0$ factor of
(\ref{eq:GammaWfactor}).  For non-Abelian gauge groups, there is, in
addition to the local term (\ref{eq:GammaWn0}) obtained in perturbation
theory, also a nonlocal term in $\tilde{A}_m(\vec{x})$ which restores the
full three-dimensional gauge invariance (\ref{eq:3dgaugetr}), not just its
infinitesimal version. This nonlocal term vanishes, however, for gauge
potentials $\tilde{A}_m(\vec{x})$ sufficiently close to zero. See Refs.
\cite{ADM85,CL89} and references therein.

The integral in (\ref{eq:GammaWn0}) can be extended over the whole of
3-space, because the gauge potentials $\tilde{A}_m$ of  (\ref{eq:Atilde}),
(\ref{eq:ball})
 vanish outside the ball $B^3$.
The gauge potentials $\tilde{A}_m$ are also $x^4$-independent. Insisting
upon translation invariance, the expression (\ref{eq:GammaWn0}) can then
be written as the following four-dimensional integral:
\begin{equation}\label{eq:GammaWAtilde}
\Gamma_{\rm \,W}^{\,n=0}\,[\tilde{A}\,]  \supseteq \,
             i \int_{\R^3} \rmd^3 x \int_{0}^{L}\rmd x^4\;
             \frac{s_0 \,(1+a)\,\pi}{L}\; \omega_{\rm \,CS}
             [\tilde{A}_1(\vec{x}),\tilde{A}_2(\vec{x}),\tilde{A}_3(\vec{x})\,]\;,
\end{equation}
with $s_0 = \pm \,1$ as defined in (\ref{eq:s0}) and parameter $a=0$ for
the simple non-Abelian gauge group considered up till now. The one-loop
calculation for three-dimen\-sio\-nal Abelian $U(1)$ gauge potentials
gives essentially the same result \cite{R84,P87}, with the factor $\pi$ in
(\ref{eq:GammaWn0}) replaced by $2\, \pi$ and the parameter $a$ in
(\ref{eq:GammaWAtilde}) set equal to $1$.

The \CS~term (\ref{eq:GammaWAtilde}) is the main result of this paper. The
result was obtained for the particular chiral gauge theory with the gauge
group $G$ $=$ $SO(10)$ and the fermion representation $R_{\,L}$ $=$
$\mathbf{16}$, but holds for an arbitrary simple compact Lie group (or
Abelian $U(1)$ group) and an arbitrary nonsinglet irreducible fermion
representation, as long as the fermion representation is normalized
appropriately and the complete theory is free of chiral gauge anomalies
(see Section 5). In the next two sections, we will take a closer look at
this result and present some generalizations.

\section{Lorentz and CPT noninvariance}

For the special gauge potentials (\ref{eq:Atilde}), (\ref{eq:ball}) and
the Euclidean spacetime manifold $M$ $=$ $\R^3 \times S^1$, we have found
in the previous section the emergence of a \CS~term
(\ref{eq:GammaWAtilde}) in the effective gauge field action. The
calculation, which relies on earlier results for the three-dimensional
parity anomaly, applies to both Abelian and non-Abelian gauge groups,
provided the chiral gauge anomalies cancel in the complete theory (see
Section 5).

For arbitrary gauge potentials $A_{\mu}(\vec{x},x^4)$ which drop to zero  %%
faster than $r^{-1}$ as $r \equiv |\vec{x}|$ $\rightarrow$ $\infty$ and
which have trivial holonomies (see below), the effective action term
(\ref{eq:GammaWAtilde}) can be written as the following local expression:
\begin{equation}\label{eq:GammaA}
\Gamma_{\rm \,CS-like}^{\,\R^3 \times S^1}[A\,] =  i
 \int_{\R^3} \rmd^3 x \int_{0}^{L} \rmd x^4\; \frac{s_0\,(1+a)\,\pi}{L}\,
 \omega_{\rm\,CS} [A_1(\vec{x},x^4),A_2(\vec{x},x^4),A_3(\vec{x},x^4)\,],
\end{equation}
with the \CS~density $\omega_{\rm CS}$ given by (\ref{eq:omegaCS}), an
integer factor $s_0 = \pm \,1$ defined in (\ref{eq:s0}), and an integer
parameter $a=0$ or 1 for a simple non-Abelian gauge group or an Abelian
$U(1)$ gauge group, respectively. Eq. (\ref{eq:GammaA}), for simple
non-Abelian gauge groups, is precisely equal to the counterterm presented
in Ref. \cite{K98}. The expression (\ref{eq:GammaA}) is called \CS-like,
because a genuine topological \CS~term exists only in an odd number of
dimensions \cite{CS74}. Remark that this \CS-like term (\ref{eq:GammaA})
comes from a combination of infrared ($1/L$) and ultraviolet ($s_0$
$\equiv$ $\Lambda_0/|\Lambda_0|$) effects.

The local \CS-like term (\ref{eq:GammaA}) has the important property of
invariance under \emph{infinitesimal} four-dimensional gauge
transformations. (This property would not hold if the particular
\CS~density $\omega_{\rm CS}$ as given by (\ref{eq:GammaA}) were replaced
by, for example, $\omega_{\rm CS}
[\mathcal{A}_1(\vec{x}),\mathcal{A}_2(\vec{x}),\mathcal{A}_3(\vec{x})\,]$,
with the averaged gauge potentials $\mathcal{A}_m(\vec{x})$ $\equiv$
$L^{-1}\int_{0}^{L} \rmd x^4$ $A_m(\vec{x},x^4)$. Of course, such an
effective action term using the averaged gauge potentials $\mathcal{A}_m$
would not be local either.) For simple compact connected Lie groups $G$,
there are also \emph{large} gauge transformations with a gauge function
$g$ $=$ $g(\vec{x})$ $\in$ $G$ corresponding to a nontrivial element of
the homotopy group $\pi_3(G)$ $=$ $\Z$. As mentioned in Section 3, there
is a nonlocal term in the effective gauge field action which restores
invariance under these finite gauge transformations, but this nonlocal
term vanishes for gauge potentials $A_m(\vec{x},x^4)$ sufficiently close
to zero. In addition, there are, for Lie groups $G$ $=$ $SO(N \geq 3)$ or
$U(1)$ with homotopy group $\pi_1(G)$ $\neq$ $0$, large gauge
transformations with gauge function $g$ $=$ $g(x^4)$ $\in$ $G$, but the
\CS-like term (\ref{eq:GammaA}) is obviously invariant under these
particular finite gauge transformations.

Turning to spacetime transformations, the effective action term
(\ref{eq:GammaA}) is clearly invariant under translations. The \CS~density
in its integrand, though, involves only three of the four gauge potentials
$A_\mu(x)$ and three of the six components of the field strength tensor
$A_{\mu\nu}(x)$ $\equiv$ $\partial_\mu A_\nu$ $-$ $\partial_\nu A_\mu$ $+$
$\left[\,A_\mu\, ,A_\nu\,\right]$, which makes the effective gauge field
action manifestly Lorentz noninvariant. Physically, this would, for
example, lead to anisotropic propagation (birefringence) of the gauge
boson fields \cite{CFJ90}.

We are not able to determine the imaginary part of the effective gauge
field action exactly. (The effective action might, for example, have some
dependence on the trace of the path-ordered exponential integral
(holonomy) ${\rm tr}\,h(\vec{x})$ $\equiv$ ${\rm
tr\;P}\,\exp\,\{\int_{0}^{L}\rmd x^4 \,A_4(\vec{x},x^4)\}$, which could
not have been detected by the gauge potentials (\ref{eq:Atilde}) used in
Section 3. The effective action term (\ref{eq:GammaA}) holds, most likely,
only for trivial holonomies $h(\vec{x})$ $=$ $\id\,$.) But the partial
result (\ref{eq:GammaWAtilde}) suffices for the main purpose of this
paper. The appropriate CPT transformation \cite{L57,IZ80} for an
anti-Hermitian gauge potential is, namely,
\begin{equation}\label{eq:ACPT}
A_\mu(x) \, \rightarrow \, A_\mu^{\rm T}(-x)\; ,
\end{equation}
with the suffix T indicating the transpose of the matrix. For a Hermitian
electromagnetic vector potential $a_\mu(x)$, this corresponds to the usual
transformation $a_\mu(x)$ $\rightarrow$ $-a_\mu(-x)$. Using
(\ref{eq:ACPT}), one then readily verifies that the \YM~action density
${\rm tr}\left( A_{\mu\nu}A^{\mu\nu}\right)$ is CPT-even and that the
integrand of the effective action term (\ref{eq:GammaWAtilde}), or
(\ref{eq:GammaA}) for that matter, is CPT-odd. (The overall factor $i$ in
(\ref{eq:GammaWAtilde}), or (\ref{eq:GammaA}), is absent for spacetime
metrics with Lorentzian signature and need not be complex conjugated.)
This establishes the CPT anomaly for chiral gauge theories with
left-handed fermions in an arbitrary nonsinglet irreducible representation
(provided the chiral gauge anomalies cancel in the complete theory) and
spacetime manifold $M$ $=$ $\R^3\times S^1$.

The Abelian Chern-Simons density has no cubic term and the integrand of
the Abelian version of (\ref{eq:GammaA}) is odd under both CPT and T (and
even under both C and P), provided  $x^4$ corresponds to a spatial
coordinate after the Wick rotation from Euclidean to Lorentzian metric
signature. The T and CPT violation would, for example, show up in the
anisotropic propagation of the circular polarization modes of these
Abelian gauge fields (a given circular polarization mode would,
generically, have different phase velocity for propagation in opposite
directions \cite{CFJ90}).

Note, finally, that the terms (\ref{eq:GammaA}) applied to the gauge
groups $SU(3)$, $SU(2)$, and $U(1)$, with undetermined coefficients
replacing $s_0 \,(1+a)\, \pi/L$, have also been considered in a
Standard-Model extension with Lorentz and CPT violation \cite{CK97}. As
discussed above, these $SU(3)$ and $SU(2)$ \CS-like terms are noninvariant
under certain large gauge transformations \cite{K98} and the $U(1)$
\CS-like term may also become gauge dependent if magnetic flux from
monopoles is allowed for \cite{P87}. This suggests that either the
corresponding coefficients must be zero or that additional nonlocal terms
restoring gauge invariance must be included in the theory. The anomaly
calculation of the present paper follows the second path, with nonlocal
terms restoring gauge invariance.\footnote{The same is to be expected for
the effective gauge field action from fermions with \emph{explicit}
CPT-violating, but gauge-invariant, terms in the action \cite{CK97}. For
recent results on the induced Abelian \CS-like term from a massive Dirac
fermion with a CPT-violating axial-vector term in the action, see Ref.
\cite{JK99} and references therein. Chiral fermions with a real chemical
potential $\mu$ (and a corresponding CPT-odd term in the action) also give
rise to an induced \CS-like term, which is now proportional to $\mu$, see
Ref. \cite{RW85} and the last equation therein.}

\section{Generalizations}

The effective gauge field action for certain chiral gauge theories defined
over a fixed four-di\-men\-sio\-nal Euclidean spacetime manifold $M$ with
Cartesian product structure $\R^3 \times S^1$ has been found to contain a
CPT-violating term if gauge invariance is maintained. For the special
gauge potentials (\ref{eq:Atilde}), this CPT-violating term is given by
(\ref{eq:GammaWAtilde}), which can be written as (\ref{eq:GammaA}) for
arbitrary localized gauge potentials with trivial holonomies. As mentioned
in the previous section, the overall factor $i$ in (\ref{eq:GammaWAtilde})
and (\ref{eq:GammaA}) would be absent for a Lorentzian signature of the
metric.

Essentially the same result holds for other orientable spacetime manifolds
$M$, as long as at least one compact spatial dimension can be factored
out. The crucial point is that the Weyl operator should be separable with
respect to this compact coordinate. Also, the spin structure over the
compact spatial dimension must be such as to allow for zero momentum of
the fermions, cf. Eq. (\ref{eq:sumI3}). One example would be the flat
Euclidean spacetime manifold $M$ $=$ $\R^3 \times \mathrm{I}$, with the
closed in\-ter\-val $\mathrm{I}$ $\equiv$ $[0,L\,]$ $\subset$ $\R$
replacing the circle $S^1$ considered before. Here, the chiral fermions
are taken to have free \bcs~over $\mathrm{I}$. (There would be no CPT
anomaly for strictly antiperiodic \bcs. This would be the case for
finite-temperature field theory in the Euclidean path integral formulation
\cite{IZ80,RW85}, which uses the same manifold $\R^3 \times \mathrm{I}$
with antiperiodic \bcs~for the fermions over the in\-ter\-val $\mathrm{I}$
$\equiv$ $[0,\beta\,]$, where $\beta$ stands for the inverse temperature.)
Another example would be the flat Minkowski-like manifold $M$ $=$ $\R
\times S^1 \times S^1 \times S^1$, with time $t \in \R$ and a compact
space manifold, which would have three possible terms of the form
(\ref{eq:GammaA}) in the effective gauge field action. Similar effects may
occur in higher- and lower-dimensional chiral gauge theories, but for the
rest of this section we concentrate on the four-dimensional case, again
with the spacetime manifold $M$ $=$ $\R^3 \times S^1$.

The calculation of the CPT-violating term in Section 3 was performed for a
single irreducible representation of left-handed Weyl fermions. The
particular theory considered, with the gauge group $G$ $=$ $SO(10)$ and
the fermion representation $R_{\,L}$ $=$ $\mathbf{16}$, is free of chiral
gauge anomalies \cite{B69,GJ72,KR97,W82}. It is, of course, possible to
have more than one nonsinglet irreducible fermion representation $r$ for
the left-handed fermions, provided the chiral anomalies cancel. The
reducible fermion representation is then $R_{\,L}$ $=$ $\sum_{f}\,r_f$,
with the label $f$ running over $1$, $\ldots$ , $N_F$. Here, and in the
following, the gauge group $G$ is taken to be either a simple compact Lie
group or an Abelian $U(1)$ group.

Vectorlike gauge theories with, for example, one nonsinglet irreducible
representation $r$ have $R_{\,L}$ $=$ $r$ $+$ $\bar{r}$ and corresponding
three-dimensional \PV~masses $\Lambda_{0r}$ and $\Lambda_{0\bar{r}}$,
where $\bar{r}$ denotes the conjugate representation of $r$.
Four-dimensional parity invariance gives $\Lambda_{0r}$ $=$ $-
\Lambda_{0\bar{r}}\,$. Recalling (\ref{eq:s0}), this implies that the CPT
anomaly (\ref{eq:GammaA}) cancels for this particular vectorlike gauge
theory. The same cancellation occurs, in fact, for \emph{any} vectorlike
gauge theory.

Chiral gauge theories with $R_{\,L}$ $=$ $\sum_{f}\,r_f$ (and $R_{\,L}$
$\neq$ $\bar{R}_{\,L}$) may or may not have a CPT-violating term
(\ref{eq:GammaA}) left over in the effective gauge field action, depending
on the relative signs of the corresponding three-dimensional \PV~masses
$\Lambda_{0f}$. Of course, attention must be paid to the normalization of
the different irreducible representations $r_f$. Also, the situation can
be complicated further by having more three-dimensional \PV~fields than
the ones used in Section 3. The factor $s_0=\pm\,1$ in
(\ref{eq:GammaWn0}), (\ref{eq:GammaWAtilde}), and (\ref{eq:GammaA}), is
then replaced by $(2\,k_{0f}+1)$, for $k_{0f} \in \Z$. The same odd
integer prefactors of the induced \CS~density also appear for other
three-dimensional ultraviolet regularization methods and are, in fact, to
be expected on general grounds \cite{CL89}.

For a \cgth~with an \emph{odd} number $N_F$ of \emph{equal} irreducible
left-handed fermion representations, there necessarily appears a
CPT-violating term proportional to (\ref{eq:GammaA}) in the effective
gauge field action. The reason is that the sum of an odd number of odd
numbers does not vanish, $\sum_f\,(2\,k_{0f}+1)$ $\neq$ $0$  for $f$
summed over $1$ to $N_F$. An example for $N_F=3$ would be the $SO(10)$
chiral gauge theory with the reducible fermion representation $R_{\,L}$
$=$ $\mathbf{16}$ $+$ $\mathbf{16}$ $+$ $\mathbf{16}$, which necessarily
has a CPT-violating term proportional to (\ref{eq:GammaA}) for the
$SO(10)$ gauge fields in the effective action.

This particular $SO(10)$ model contains, as is well known, the $SU(3)$
$\times$ $SU(2)$ $\times$ $U(1)$ \SM~\cite{IZ80,Z82} with $N_F=3$ families
of 15 left-handed Weyl fermions (quarks and leptons), together with
$N_F=3$ left-handed Weyl fermion singlets (conjugates of the hypothetical
right-handed neutrinos). The \SM~has thus a CPT-violating term
proportional to (\ref{eq:GammaA}) for the hypercharge $U(1)$ gauge fields
in the effective action, together with similar terms for the weak $SU(2)$
and color $SU(3)$ gauge fields, as long as the spacetime manifold has the
appropriate Cartesian product structure and the \SM~is embedded in this
particular $SO(10)$ chiral gauge theory.\footnote{It is not clear to what
extent the 33 remaining gauge bosons from $SO(10)$ need to be physical,
but they could always be given large masses by the Higgs mechanism
\cite{GG74,Z82}.} Other simple compact Lie groups instead of $SO(10)$ may
also be used for the embedding of the \SM~fermions, as long as they have
an odd number of equal irreducible representations. This embedding
condition for the \SM~fermions \emph{guarantees} the presence of the
CPT-violating \CS-like terms for the $SU(3)$ $\times$ $SU(2)$ $\times$
$U(1)$ gauge fields in the effective action, otherwise these terms may or
may not appear, depending on the regularization scheme.

\section{Discussion}

In the previous sections, we have established for certain chiral gauge
theories defined over the spacetime manifold $M$ $=$ $\R^3 \times S^1$ the
necessary presence of a CPT-violating term in the gauge-invariant
effective action. The question, now, is what happened to the CPT theorem?
It appears that the CPT theorem is circumvented by a breakdown of local
Lorentz invariance at the quantum level. (See also the paragraph above
(\ref{eq:GammaWproduct}), which discusses the breaking of Lorentz
invariance by the ultraviolet regularization used.) More specifically, the
second-quantized vacuum seems to play a role in connecting the global
spacetime structure to the local physics. The next two paragraphs
elaborate this point but may be skipped in a first reading.

For non-Abelian chiral gauge groups, there is the condition of gauge
invariance to deal with in these particular \qfths~which are potentially
afflicted by the nonperturbative chiral gauge anomaly discovered earlier
\cite{KR97}. This nonperturbative chiral gauge anomaly depends on the
global spacetime structure in a Lorentz noninvariant way, one spatial
direction being singled out by the so-called $Z$-string configuration
responsible for the gauge anomaly in the Hamiltonian formulation. If the
theory has indeed this $Z$-string chiral gauge anomaly, then the
restoration of gauge invariance obviously requires interactions which are
themselves Lorentz noninvariant \cite{K98}. But, even if the theory does
not have a net $Z$-string chiral gauge anomaly, there still occurs, in
first-quantization, the spectral flow which treats one spatial dimension
differently from the others \cite{KR97}. This implies that a tentative
second-quantized vacuum state varies along the corresponding loop of gauge
transformations. Imposing gauge invariance throughout then leads to
Lorentz noninvariance of the theory. In both cases, the invariance under
the proper orthochronous Lorentz group is lost and the CPT theorem no
longer applies \cite{L57}. It is then possible to have a non-Abelian
CPT-odd term (\ref{eq:GammaA}) in the effective gauge field action.

For Abelian chiral gauge groups, nonzero magnetic flux from monopoles can
also give rise to spectral flow, as discussed for the three-dimensional
case in Ref. \cite{P87}. For the four-dimensional case, this setup again
breaks Lorentz invariance (just as the $Z$-string does for the non-Abelian
chiral gauge anomaly) and the CPT theorem no longer applies, with the
possibility of having an Abelian  CPT-odd term (\ref{eq:GammaA}) in the
effective gauge field action.

For both Abelian and non-Abelian chiral gauge groups, it remains to be
seen whether or not the gauge-invariant, but CPT-violating, theory is
consistent. In particular, the properties of microcausality and positivity
of the energy need to be established, cf. Refs. \cite{S64,SW64,CK97}. If
the theory in question turns out to be inconsistent, then perhaps it could
be interpreted as part of a more fundamental theory, possibly involving
curvature and torsion.

In this paper, we have primarily been concerned with the mechanism of the
CPT anomaly, not potential applications. Let us, however, mention two
possibilities. First, there may be the ``optical activity'' \cite{CFJ90}
discussed in the Introduction, where the linear polarization of a plane
wave of gauge fields gets rotated \emph{in vacuo} (in our case, through
the quantum effects of the chiral fermions encoded in the effective gauge
field action). As mentioned in Section 5, the phenomenon could occur for
the photon field of the \SM, as long as there is the $SO(10)$-like
embedding of the \SM~fermions and the appropriate Cartesian product
structure of the spacetime manifold. The laboratory measurement of this
optical activity of the vacuum could, in principle, provide information
about the global structure and size of the universe. More realistically,
the mass scale of the CPT-violating term (\ref{eq:GammaA}) for the photon
field is of the order of
\begin{equation}\label{eq:mass-scale}
\alpha\, L^{-1} \sim \, 10^{-35}\, {\rm eV}\, \left( \frac{\alpha}{1/137}
\right)\, \left( \frac{1.5\;10^{10}\,{\rm lyr}}{L} \right)\; ,
\end{equation}
with $\alpha$ $\equiv$ $e^2/4\,\pi$ the fine-structure constant and $L$
the range of the compact spatial coordinate. This mass is, of course, very
small on the scale of the known elementary particles (the present universe
being very large), but, remarkably, it is only a factor $100$ below the
current upper bound of $\sim \! 10^{-33}\, {\rm eV}$ obtained from
observations on distant radio galaxies, see Refs. \cite{CFJ90,WPC97} and
references therein. (The ``laboratory'' has now been expanded to a
significant part of the visible universe.) A de\-di\-ca\-ted observation
program to map the linear polarization in a large number of distant radio
sources \cite{WPC97}, or future satellite experiments to measure the
polarization of the cosmic microwave background \cite{CCT94}, can perhaps
reach the sensitivity level set by (\ref{eq:mass-scale}).

Second, the CPT anomaly may have been important in the very early
universe. In the present paper, we have considered a fixed spacetime
manifold with given topology. With gravity, spacetime becomes dynamic. For
an inverse size $1/L(t)$ and  typical scattering energies of the order of
the gravitational scale (Planck mass), the CPT-violating effects of the
effective action term (\ref{eq:GammaA}) are relatively unsuppressed
compared to gravity, that is suppressed by the square of the gauge
coupling constant only. Of course, the fundamental theory of gravity
remains to be determined if there is indeed Lorentz noninvariance in
certain inertial frames. Still, it is conceivable that the CPT anomaly
plays a role in defining a ``fundamental arrow-of-time'' \cite{W80,P79},
as a quantum mechanical effect coming from the interplay of chiral
fermions, gauge field interactions and the topology of spacetime.

\newpage
\section*{Acknowledgements} This work was completed at the Niels Bohr Institute
and we gratefully acknowledge the hospitality of the High-Energy Theory
group. We also thank C. Adam and V.A. Kosteleck\'{y} for comments on the
manuscript and R. Jackiw and the referee for pointing out some additional
references.


\begin{thebibliography}{99}
\bibitem{L57}    G. L\"{u}ders,  Ann. Phys. (N. Y.) 2 (1957) 1.
\bibitem{S64}    J. Sakurai, Invariance Principles and Elementary
                 Particles, Princeton Univ. Press, Princeton, 1964.
\bibitem{SW64}   R. Streater, A. Wightman, PCT, Spin and Statistics,
                 and All That, Benjamin, New York, 1964.
\bibitem{W80}    R. Wald, Phys. Rev. D 21 (1980) 2742.
\bibitem{H82}    S. Hawking, Commun. Math. Phys. 87 (1982) 395;
                 Phys. Rev. D 32 (1985) 2489.
\bibitem{KP91}   V. Kosteleck\'{y}, R. Potting, Nucl. Phys. B 359 (1991) 545.
\bibitem{K98}    F. Klinkhamer, Nucl. Phys. B 535 (1998) 233.
\bibitem{YM54}   C. Yang, R. Mills, Phys. Rev. 96 (1954) 191.
\bibitem{W29}    H. Weyl, Z. Phys. 56 (1929) 330.
\bibitem{A69}    S. Adler, Phys. Rev. 177 (1969) 2426.
\bibitem{BJ69}   J. Bell, R. Jackiw, Nuovo Cimento A 60 (1969) 47.
\bibitem{B69}    W. Bardeen, Phys. Rev. 184 (1969) 1848.
\bibitem{GJ72}   D. Gross, R. Jackiw, Phys. Rev. D 6 (1972) 477.
\bibitem{WZ71}   J. Wess, B. Zumino, Phys. Lett. B 37 (1971) 95.
\bibitem{W83}    E. Witten, Nucl. Phys. B 223 (1983) 422.
\bibitem{KR97}   F. Klinkhamer, C. Rupp, Nucl. Phys. B 495 (1997) 172.
\bibitem{W82}    E. Witten, Phys. Lett. B 117  (1982) 324.
\bibitem{R84}    A. Redlich, Phys. Rev. Lett. 52 (1984) 18;
                 Phys. Rev. D 29 (1984) 2366.
\bibitem{NS83}   A. Niemi, G. Semenoff, Phys. Rev. Lett. 51 (1983) 2077.
\bibitem{AW84}   L. Alvarez-Gaum{\'e}, E. Witten, Nucl. Phys. B 234 (1984) 269.
\bibitem{ADM85}  L. Alvarez-Gaum{\'e}, S. Della Pietra, G. Moore,
                 Ann. Phys. (N. Y.) 163 (1985) 288.
\bibitem{P87}    A. Polychronakos, Nucl. Phys. B 281 (1987) 241.
\bibitem{CL89}   A. Coste, M. L\"{u}scher, Nucl. Phys. B 323 (1989) 631.
\bibitem{CS74}   S. Chern, J. Simons, Ann. Math. 99 (1974) 48.
\bibitem{CFJ90}  S. Carroll, G. Field, R. Jackiw, Phys. Rev. D 41 (1990) 1231.
\bibitem{IZ80}   C. Itzykson, J.-B. Zuber, Quantum Field Theory,
                 McGraw-Hill, New York, 1980.
\bibitem{GG74}   H. Georgi, S. Glashow, Phys. Rev. Lett. 32 (1974) 438.
\bibitem{Z82}    A. Zee (Ed.),
                 Unity of Forces in the Universe, World Scientific, Singapore, 1982.
\bibitem{FS86}   L. Faddeev, S. Shatashvili, Phys. Lett. B 167 (1986) 225.
\bibitem{N98}    H. Neuberger, Phys. Lett. B 417 (1998) 141;
                 Phys. Lett. B 427 (1998) 353.
\bibitem{L99}    M. L\"{u}scher, Nucl. Phys. B 549 (1999) 295;
                 Nucl. Phys. B 568 (2000) 162.
\bibitem{P58}    W. Pauli, Theory of Relativity, Pergamon Press, Oxford, 1958.
\bibitem{PV49}   W. Pauli, F. Villars, Phys. Rev. 21 (1949) 434.
\bibitem{CK97}   D. Colladay, V. Kosteleck\'{y}, Phys. Rev. D 55 (1997) 6760;
                 D 58 (1998) 116002.
\bibitem{JK99}   R. Jackiw, V. Kosteleck\'{y}, Phys. Rev. Lett. 82 (1999) 3572.
\bibitem{RW85}   A. Redlich, L. Wijewardhana, Phys. Rev. Lett. 54 (1985) 970.
\bibitem{WPC97}  J. Wardle, R. Perley, M. Cohen, Phys. Rev. Lett. 79 (1997) 1801.
\bibitem{CCT94}  D. Coulson, R. Crittenden, N. Turok, Phys. Rev. Lett. 73 (1994) 2390.
\bibitem{P79}    R. Penrose, in: S. Hawking, W. Israel (Eds.),
                 General Relativity: An Einstein Centenary Survey,
                 Cambridge Univ. Press, Cambridge, 1979, Section 12.4.
\end{thebibliography}
\end{document}